\documentclass[twocolumn,aip,jmp]{revtex4}

\usepackage{graphicx}
\usepackage{dcolumn}
\usepackage{bm}
\usepackage{amssymb}
\usepackage{amsmath}
\usepackage{wasysym}
\usepackage{color}

\usepackage{placeins}
\usepackage{epstopdf}
\begin{document}

\preprint{APS}

\title[Kurztitel]{Quantized electronic fine structure with large anisotropy in ferromagnetic Fe films}

\author{C. Seibel$^{1}$, A. Nuber$^{1}$, H. Bentmann$^{1}$, M. Mulazzi$^{1}$, P. Blaha$^2$, G. Sangiovanni$^3$, F. Reinert$^{1}$}

\affiliation {$^1$Experimentelle Physik VII, Universit\"at W\"urzburg, Am Hubland, 97074 W\"urzburg, Germany\\
$^2$Institute of Materials Chemistry, TU Vienna, 1060 Vienna, Austria\\
$^3$Institute for Theoretical Physics and Astrophysics, University of W\"urzburg, 97074 W\"urzburg, Germany}
\date{\today}

\begin{abstract}

We report on the spectroscopic observation of a quantized electronic fine structure near the Fermi energy in thin Fe films grown on W(110). The quantum well states are detected down to binding energies of $\sim$10~meV by angle-resolved photoelectron spectroscopy. The band dispersion of these states is found to feature a pronounced anisotropy within the surface plane: It is free-electron like along the $\overline{\Gamma \rm{H}}$-direction while it becomes heavy along $\overline{\Gamma \rm{N}}$. Density functional theory calculations identify the observed states to have both majority and minority spin character and indicate that the large anisotropy can be dependent on the number of Fe-layers and coupling to the substrate. 
\end{abstract}

\maketitle

The physical and chemical properties of a solid state system whose dimensions are on the nanometer scale may drastically differ from those of real bulk materials. This is because interfaces or surfaces become increasingly relevant and thus gives rise to quantum confinement, reduced atomic coordination numbers and a broken crystal symmetry. As a result, low-dimensional structures such as clusters \cite{Cox:93} or quantum wells \cite{Chiang:00,wiessner:12} offer the opportunity to modify magnetic, electronic or lattice properties with atomic-scale precision or to even induce entirely new effects \cite{Molenkamp:07,Bode:07}. In ultrathin films the size reduction to a few nanometer thickness leads to the formation of discrete quantum-well states (QWS) \cite{Chiang:00,Smith:94.1}. The resulting thickness-dependence of the electronic structure has been shown to affect various fundamental physical mechanisms, such as superconductivity \cite{Xue:04}, electron-phonon coupling \cite{Chiang:02.7,Xue:05}, spin-orbit coupling \cite{Bentmann:12.05}, or magnetocrystalline anisotropy \cite{Wu:09.5}.

Angle-resolved photoelectron spectroscopy (ARPES) is a powerful technique to probe the momentum-dependent electronic structure of solids and has been successfully employed to determine the properties of QWS in thin film systems (see e.g. Refs.~\cite{Chiang:00,Smith:94.1,speer:03.06,forster2011}). Furthermore, with high-resolution ARPES it is feasible to detect narrow spectral features near the Fermi energy on the scale of a few meV which has allowed for spectroscopic investigations of superconducting gaps \cite{reinert00} or heavy-fermion bands \cite{Laubschat:08,santander-syro:09}. 
Single particle features, however, can sometimes also appear on this energy scale \cite{Klein:08.07}. In the present work we use high-resolution ARPES to investigate the thickness-dependent low-energy electronic structure of ultrathin {\it bcc} Fe films, a material system that is of broad interest in various fields because of its ferromagnetic properties. Our measurements for Fe grown on W(110) reveal a quantized electronic fine structure with QWS exhibiting a highly anisotropic dispersion at binding energies down to $\sim$10~meV below the Fermi level. This result shows that quantization effects influence the electronic structure of thin ferromagnetic films on this low energy scale, bearing general implications for magnetism in nanostructures. 

%\begin{figure}[b]
  %\centering
  %\begin{minipage}[T!]{0.238\textwidth}
    %\includegraphics[width=1\textwidth]{fig1_a} 
    %
  %\end{minipage}
  %\begin{minipage}[T!]{0.238\textwidth}
    %\includegraphics[width=1\textwidth]{fig1_b}  
    %
  %\end{minipage}
  %\caption{(color online). (a) First-order LEED spots for 20 ML \textit{bcc}-Fe(110) on W(110) and the corresponding surface Brillouin zone with high-symmetry points. (b) ARPES band structure along $\overline{\Gamma \rm{H}}$ for the same film thickness. The spectrum shows the second derivative of the original data set. The color-scale was adjusted individually for the segments i and ii.}
  %\label{fig1}
  %\end{figure}
					
\begin{figure}[b]
\includegraphics[width=0.5\columnwidth]{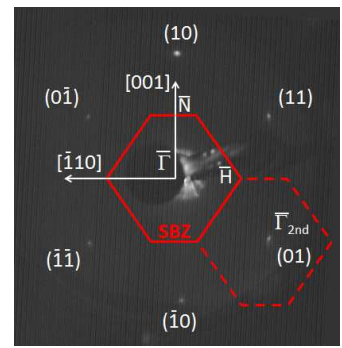}
\caption{(color online) First-order LEED spots for 20 ML \textit{bcc}-Fe(110) on W(110) measured at electron energy of $81$~eV and the surface Brillouin zone with high-symmetry points.}
\label{fig1}
\end{figure}

The presented ARPES data were acquired using a \mbox{SCIENTA R4000} hemispherical electron spectrometer and a monochromatized He discharge lamp. For all measurements we used an excitation energy of 21.22~eV (He~I$_{\alpha}$). The energy and angular resolutions were 5~meV and 0.3$^\circ$, respectively. The experiments were performed in ultra-high vacuum (UHV) at a base pressure of $1\times 10^{-10}$ mbar and at sample temperatures of 30--140~K. A clean and well-ordered W(110) surface was prepared according to the procedure described in Ref.~\onlinecite{Zakeri2010}. Fe was evaporated on the W(110) substrate at room temperature using an electron beam heated crucible at a temperature of about 1500~K. After deposition the films were annealed at 750~K. The long range order and the chemical cleanliness of the Fe films were characterized by low-energy electron diffraction (LEED), see Fig.~\ref{fig1}, and X-ray photoemission (XPS), respectively. Absolute film thicknesses were characterized by XPS measurements with an accuracy of approximately $\pm 2$~ML. Relative changes in coverage between 18 ML and 23 ML were determined with higher precision in steps of 1~ML from the binding energy evolution of the observed QWS as function of film thickness.  

\begin{figure}[t]
  \centering
  \begin{minipage}[T!]{0.23\textwidth}
    \includegraphics[width=1\textwidth]{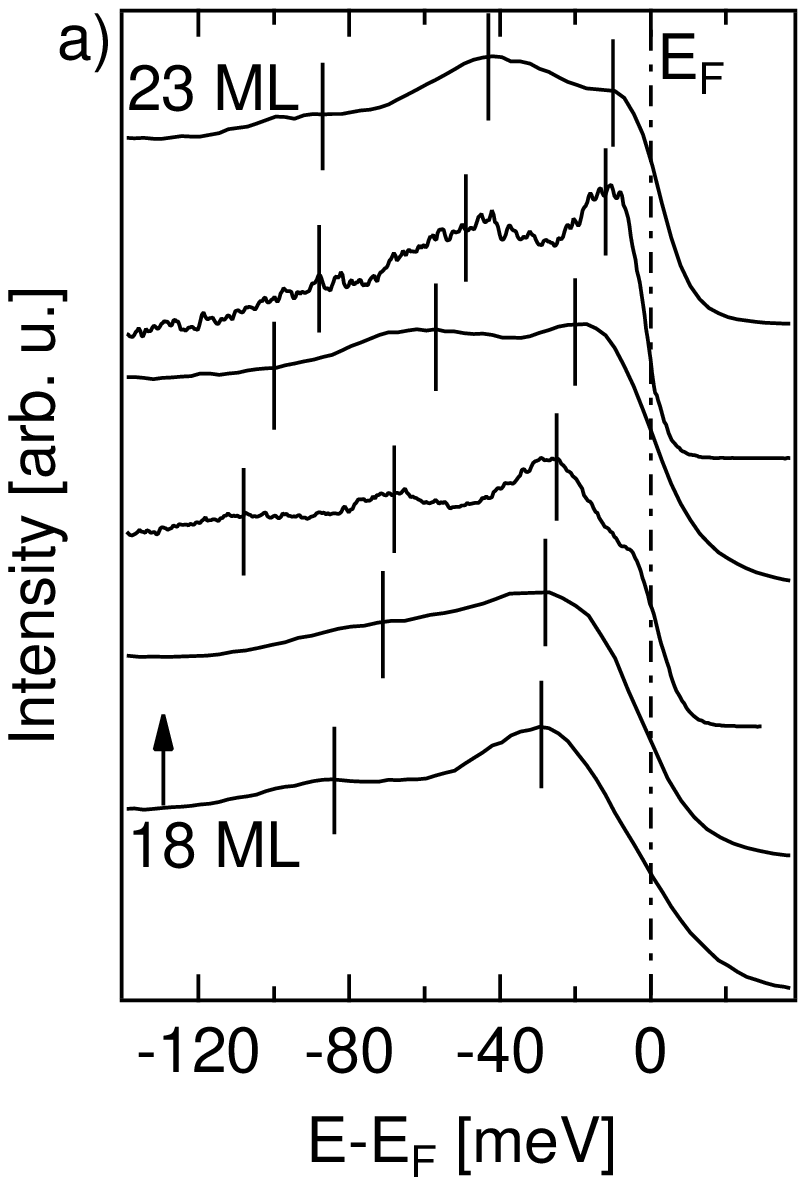} 
    
  \end{minipage}
  \begin{minipage}[T!]{0.21\textwidth}
    \includegraphics[width=1\textwidth]{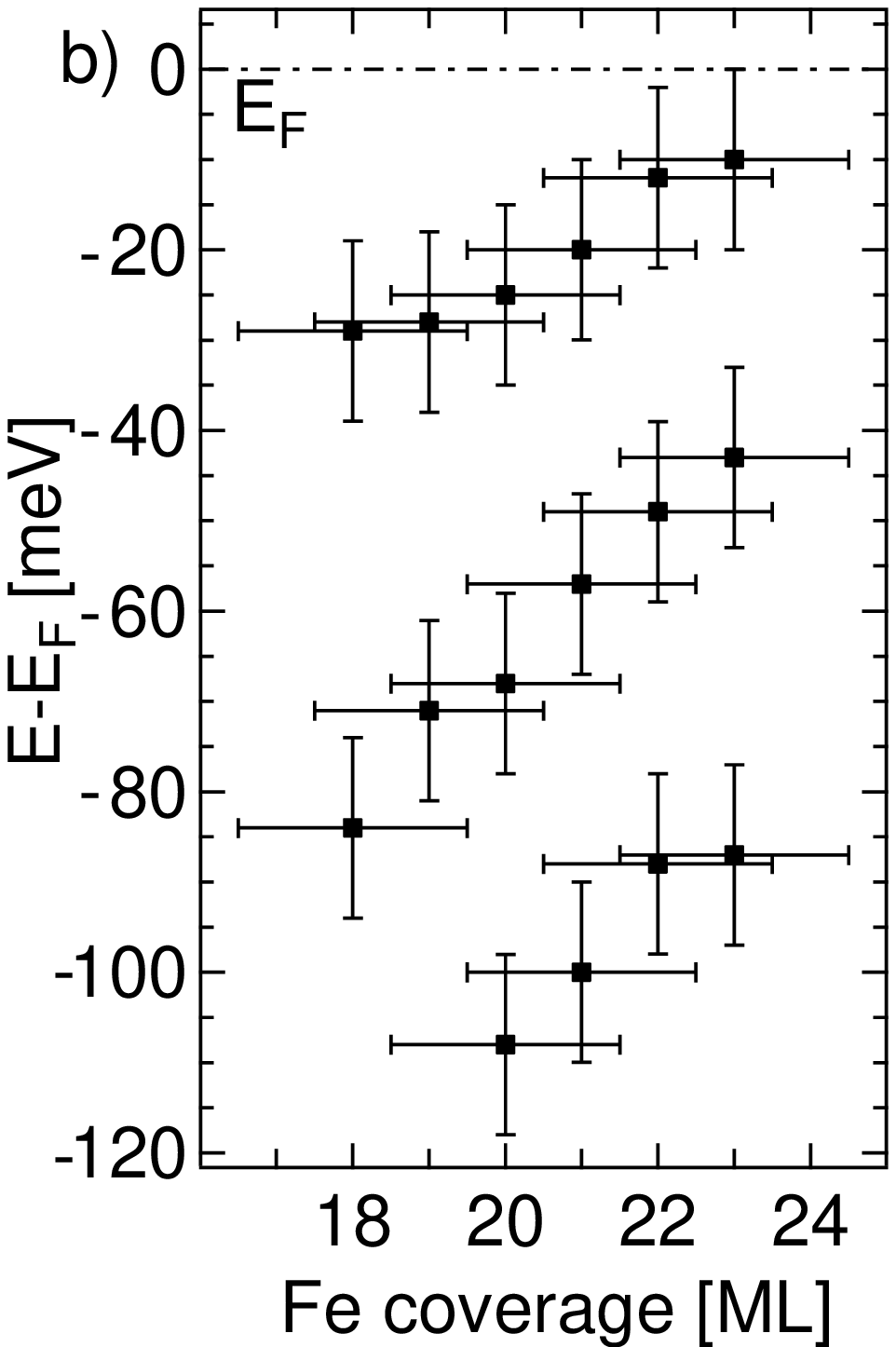}  
    
  \end{minipage}
  \caption{(a) Energy distribution curves (EDC) at the $\overline{\Gamma}$-point for Fe layers on W(110) with film thicknesses between 18~ML (bottom EDC) and 23~ML (top EDC). The spectra were obtained by integration over a k$_{||}$-range of $\pm$0.05\,\AA$^{-1}$. Peak positions corresponding to quantum well states are marked by vertical lines. (b) Binding energy evolution of the peak positions as a function of film thickness.} 
  \label{fig2}
\end{figure}

First-principles calculations were carried out using the augmented plane wave +
local orbital (APW+lo) method as implemented in the WIEN2k
code \cite{wien2k}. Exchange and correlation effects were treated within density
functional theory using the generalized gradient approximation (GGA) of Perdew et
al. \cite{pbe}. A basis set cutoff parameter $R\cdot K_{\rm{max}}=8$ (the product of
the atomic sphere radius of Fe ($R=2.15$ bohr) times the plane wave cutoff $K_{\rm{max}}$)
was used together with a 20$\times$20$\times$1 $k$--mesh. Free standing Fe(110) slabs with 20,
21~and 22~ML~Fe and a vacuum of about 40 bohr as well as one slab with (10~ML~W
/ 20~ML~Fe / vacuum) have been calculated with full geometrical relaxation.

We will start to discuss the electronic structure of Fe/W(110) near the Fermi level on the basis of the energy distribution curves (EDC) in Fig.~\ref{fig2}(a). The EDC were acquired at the $\overline{\Gamma}$-point for coverages between 18~ML and 23~ML. In all spectra we identify two or three electronic states with energy spacings of approximately 40--50~meV. Furthermore, the data provide evidence for a thickness dependence of the electronic structure.  With increasing film thickness the binding energies of all states successively shift towards the Fermi level. This electronic fine structure close to $E_F$ could not be observed in previous experiments on Fe/W(110) \cite{Schafer2007}. An overview of all measured binding energies in the investigated thickness range is given in Fig.~\ref{fig2}(b) and reveals a systematic binding energy evolution as a function of coverage. Such a behavior is a characteristic signature of quantization of electronic states along the film normal direction and we therefore identify the observed states as a sequence of QWS. 

\begin{figure}[t]
\includegraphics[width=\columnwidth]{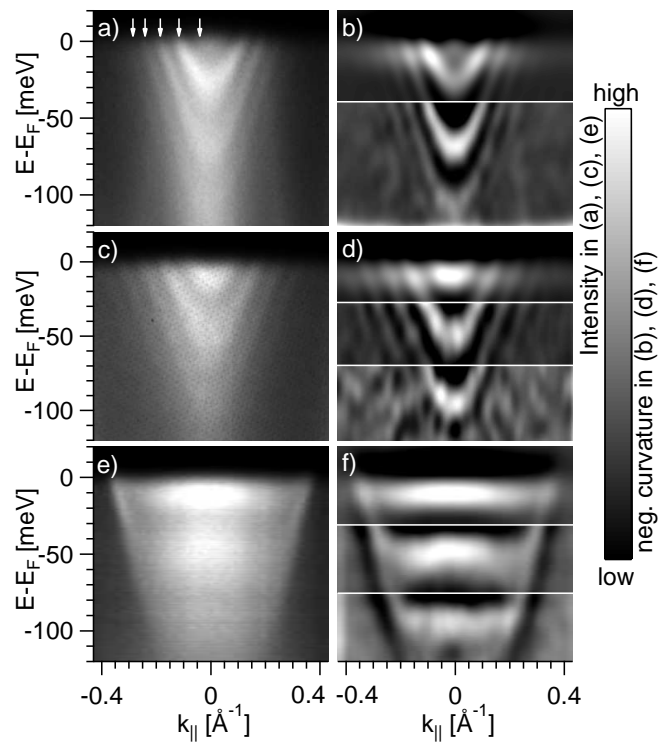}
\caption{Raw ARPES data (left) and corresponding second derivatives (right) for 20~ML Fe/W(110) along $\overline{\Gamma \rm{H}}$ in (a)-(b), for 22~ML along $\overline{\Gamma \rm{H}}$ in (c)-(d), and for 22~ML along $\overline{\Gamma \rm{N}}$ in (e)-(f). The second derivative images are divided into segments as indicated by horizontal lines. The maximum values of the greyscale were chosen individually for all of these segments in order to provide an optimal representation of all spectral features. The data evidence a free-electron-like dispersion along $\overline{\Gamma \rm{H}}$ and a flat-band behavior along $\overline{\Gamma \rm{N}}$.}
\label{fig3}
\end{figure}

In the following we address the dispersion of the QWS parallel to the film plane and its dependence on crystallographic direction. Fig.~\ref{fig3}(a) and (c) show ARPES data sets along the $\overline{\Gamma \rm{H}}$-direction for 20~ML and 22~ML, respectively. Corresponding second derivative images are presented in Fig.~\ref{fig3}(b) and (d). The second derivative procedure was performed by convoluting the raw data sets with the second derivative of a two-dimensional Gaussian with FWHM of 25~meV and 0.047~{\AA}$^{-1}$. The experimental data allow us to identify up to five QWS [arrows in Fig.~\ref{fig3}(a)] with dispersion relations that are reasonably well approximated by free electron parabolas with positive effective masses on the order of 1--2~$m_e$. Fig.~\ref{fig3}.(e) and (f) show an ARPES data set along the $\overline{\Gamma \rm{N}}$-direction and the corresponding second derivative image, respectively. It is clearly evident that along $\overline{\Gamma \rm{N}}$ the QWS do not behave as free electrons but rather form flat heavy bands. In fact, the dispersion along $\overline{\Gamma \rm{N}}$ vanishes within the experimental uncertainty over a comparably large $k$-range; the estimated lower limit for the effective mass of the QWS closest to $E_F$ is $\sim$50$m_e$. This value is comparable to heavy quasi-particle bands observed by ARPES in strongly correlated $4f$ and $5f$ compounds \cite{Laubschat:08,santander-syro:09}. It is also worth noting that the binding energy and the spectral linewidth of the state closest to $E_F$ are on the order of only $\sim$10~meV. This actually is the typical energy scale on which different many-body effects influence the electronic structure, e.g. the Kondo effect \cite{Klein:08}, superconductivity \cite{reinert00} and electron-phonon renormalization of the band mass \cite{eiguren:02.01,reinert:03.10}.   

It is well-known that electronic states in surface and thin film systems can be strongly influenced by the structural properties parallel to the interface plane \cite{Okuda:09.09,Zhong:13,Assmann:13}. A prominent example are surface states on vicinal metal surfaces that acquire a quasi-one-dimensional character due to the additional lateral confinement induced by regular arrays of step edges \cite{steppedgold,steppednickel2,steppediridium}. One may therefore suspect a similar structural origin for the large electronic anisotropy observed in the present case for Fe/W(110). However, we could rule out any effect of a possible miscut of the substrate by reproducing our experimental results on different W(110) single crystals. Furthermore, we observed sharp and isotropic LEED spots that provided no indication for the formation of a superstructure within the films, but rather gave evidence for a well-ordered (110)-surface without reconstructions (see Fig.~\ref{fig1}). We therefore identify the observed anisotropy as an intrinsic property of the Fe layers on W(110).  

\begin{figure}[t]
  \centering
  \begin{minipage}[T!]{0.238\textwidth}
    \includegraphics[width=1\textwidth]{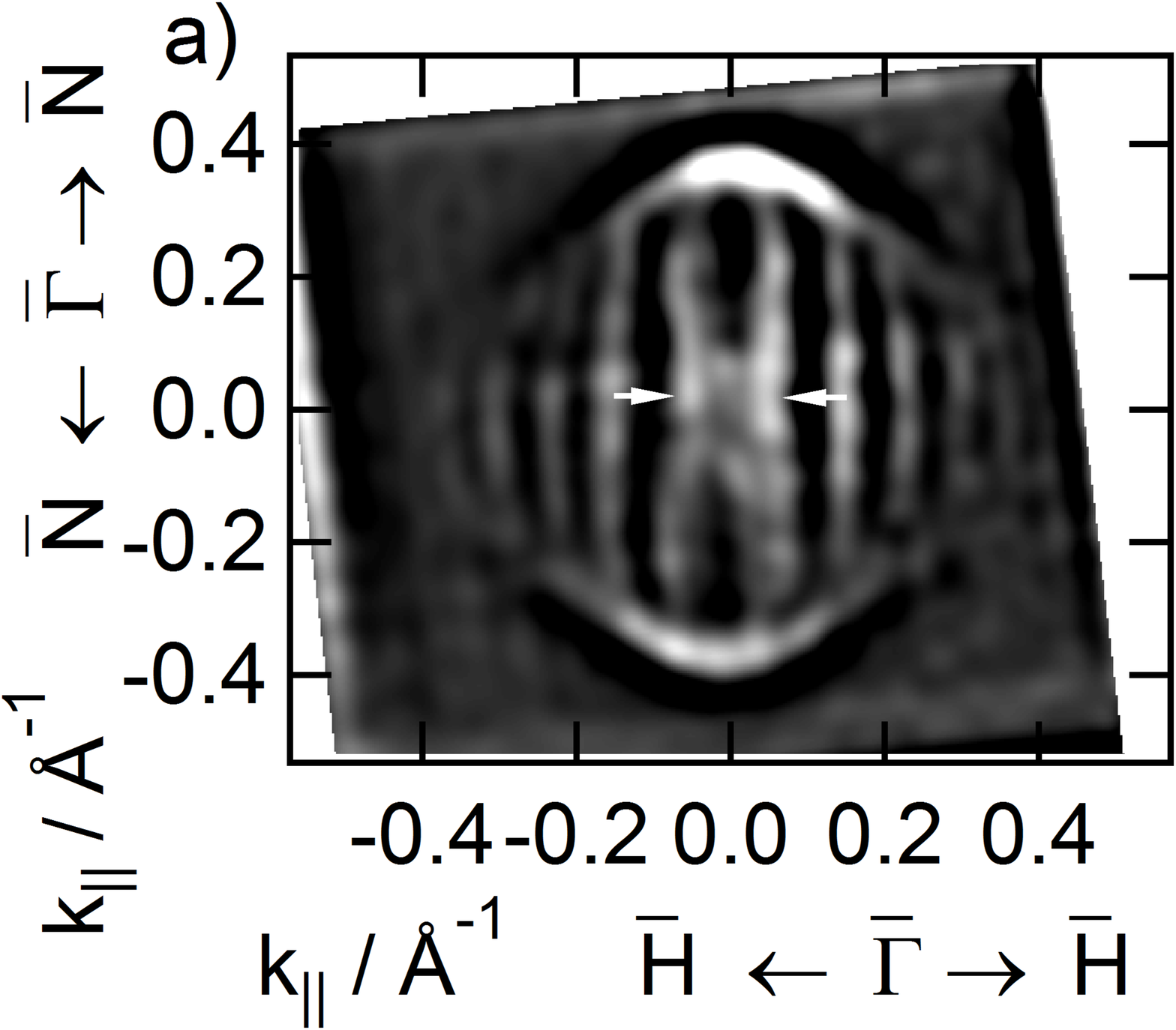} 
    
  \end{minipage}
  \begin{minipage}[T!]{0.238\textwidth}
    \includegraphics[width=1\textwidth]{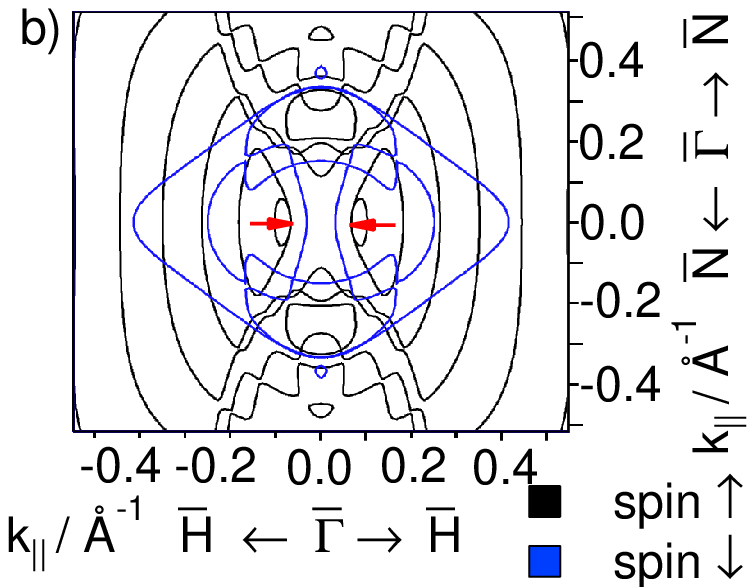}  
    
  \end{minipage}
  \caption{(color online). (a) Fermi surface for 22~ML Fe/W(110) obtained by ARPES experiments. The spectrum shows the second derivative of the ARPES data set. The experimental result is compared to the calculated Fermi surface of a free-standing 20~ML Fe slab in (b). The calculated bands are exchange split into minority (black) and majority (blue) states.} 
  \label{fig4}
\end{figure}

We will now address mani\-festa\-tions of the large aniso\-tropy in the Fermi
surface. Fig.~\ref{fig4}.(a) shows the experimental Fermi surface close to the
$\overline{\Gamma}$-point for 22~ML Fe/W(110) as determined by ARPES. The data
show the presence of a diamond-shaped outer structure and an additional inner
fine structure which could not be observed in previous ARPES experiments
\cite{Schafer2007}. The fine structure consists of parallel, almost straight
lines along $\overline{\Gamma \rm{N}}$ in accordance with the vanishing
dispersion observed for this direction. The experimental results are compared
to the calculated Fermi surface of a free-standing Fe layer displayed in
Fig.~\ref{fig4}.(b). It is evident that the very characteristic fine structure
visible in our ARPES data is remarkably well reproduced by the GGA calculation:
The main diamond-shaped structure is indeed clearly present in the
minority-spin GGA data and the characteristic lines parallel to
$\overline{\Gamma \rm{N}}$ with a weak convex deformation also show up in the
theory. Even the slight concave inwards bending of the two inner features
parallel to $\overline{\Gamma \rm{N}}$ are qualitatively reflected in the
calculation (arrows in Fig.~\ref{fig4}). Let us note that the experimental spectral weight at the Fermi surface is suppressed outside the diamond structure whereas the GGA calculation does not give a clearcut difference between the states inside and outside the diamond shape. Furthermore, the best match to the
experimental data was achieved for a film thickness of 20~ML, slightly
deviating from the experimental value of 22~ML. This difference possibly originates from the absence of a substrate in the
calculation of the Fermi surface in  Fig.~\ref{fig4}(b). The inclusion of the
substrate in the theoretical calculation is 
feasible, though only when assuming a commensurate lattice parameter for W
and Fe. In addition, the presence of a substrate makes the comparison with experiment rather difficult,
since the calculated Fermi surface contains substrate contributions that are suppressed in the experimental data due to the small electron escape depth.

\begin{figure}[t]
\includegraphics[width=\columnwidth]{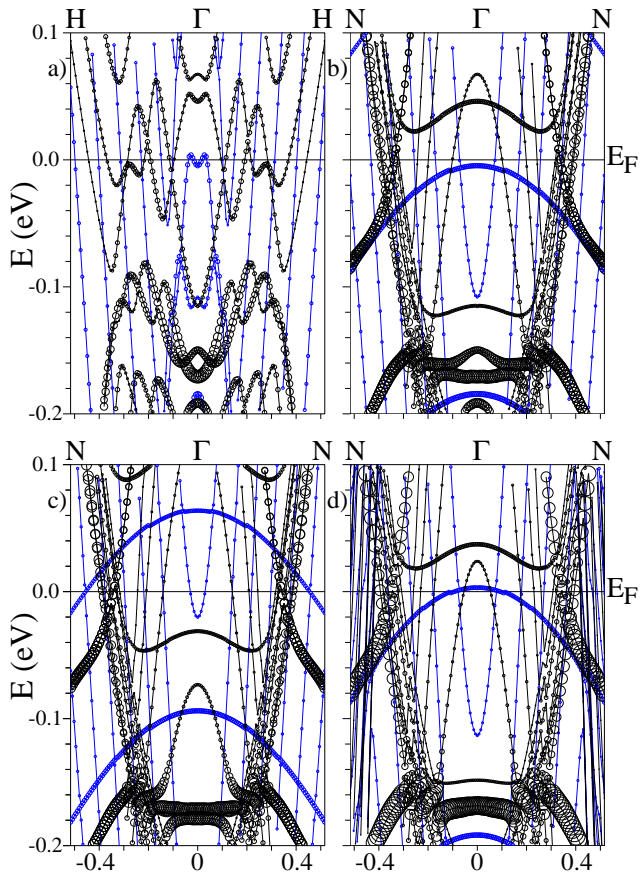} 
\caption{(color online). GGA band structures for a free-standing 20~ML Fe(110) film along the $\overline{\Gamma
    \rm{H}}$-direction in (a), for a 20~ML film along $\overline{\Gamma
    \rm{N}}$ in (b), for a 22~ML film along $\overline{\Gamma
    \rm{N}}$ in (c), and for a 20~ML film along $\overline{\Gamma
    \rm{N}}$ on W in (d). The size of
  the circles indicates the amount of Fe surface and sub-surface character, the
  horizontal axis is labeled in {\AA}$^{-1}$.
}
\label{fig5}
\end{figure}

Even if the GGA-Fermi surface map without inclusion of substrate matches the measured one in even details, a comparison of the calculated and
measured band structure is more difficult. The calculated band structure along
the $\overline{\Gamma \rm{H}}$-direction yields a
large number of parabolic bands in reasonable agreement with experiment
(Fig.~\ref{fig5}.(a)). However, the position of the main features at
$\overline{\Gamma}$ varies a lot with the number of layers or with the
inclusion of the W substrate. This can be seen more clearly
in Figs.~\ref{fig5}.(b), (c) and (d), where we show the band structure along the
$\overline{\Gamma \rm{N}}$-direction for 20 ML Fe, 22 ML Fe and 20 ML Fe/W. Apparently, the
strong, nearly parabolic feature is not much affected by the
number of layers in the calculations. On the other hand, the position of the
flat bands is extremely sensitive to the structural details. We observe a
sequence of weakly dispersive parabolic majority bands with negative curvature
in Fig.~\ref{fig5} (three such (blue) bands
for 20-layers, but only two bands are visible in this energy range for 22-layers),
whose position as well as whose
separation depend on the layer thickness. A similar situation
occurs for the fairly dispersionless minority bands (black).
Thus, the GGA for a free-standing film (or even grown on a “compressed” W substrate) yields a qualitatively correct picture of the bands, but fails to fully reproduce their exact energy position and spacing.
Unfortunately, it is almost impossible to simulate  the true change in the atomic 
structure across the interface with present computer resources. W has a much larger lattice parameter than Fe which has to induce a complicated dislocation pattern leading to the incommensurate structure observed experimentally and could only be simulated using very large supercells. For this reason we cannot get a quantitative agreement of the whole
band structure. The sensitivity of the GGA band structure to the number of layers, to the presence of the W(110) substrate and to the underlying commensurate lattice parameter is indeed a clear indication that the substrate plays a role in determining the details of the experimentally observed electronic structure.  

The above analysis points to a substantial interaction between the QWS in the Fe film with the W(110) substrate. Microscopically this interaction is caused by the hybridization of film and substrate states giving rise to a considerable extension of the QWS wave functions into the substrate \cite{speer:03.06,huang2007,forster2011}. The wave functions are therefore partially exposed to the potential landscape of the substrate which apparently promotes a further enhancement of the anisotropy in the dispersion. A recent ARPES investigation reports a highly anisotropic dispersion for a surface state on bare W(110) that, similar to the QWS in the present study, shows a steep (flat) dispersion along the $\overline{\Gamma \rm{N}}$-direction ($\overline{\Gamma \rm{H}}$-direction) \cite{Miyamoto:10.12}. The W(110) surface thus indeed appears to be prone to induce pronounced electronic anisotropies.

To the best of our knowledge no previous work has reported a comparably large electronic anisotropy for a homogeneous film so far. The present observations for Fe/W(110) are comparable to findings for nanowire reconstructions on semiconducting substrates or stepped metal surfaces with geometric structures exhibiting a one-dimensional character \cite{steppedgold,Crain:04}. The simplest approximation for a 2DEG yields an isotropic dispersion characterized by an effective mass $m^*$. Deviations from this basic scenario in real systems are more or less severe depending on the symmetry of the crystal lattice and the orbital character of the involved states. Comparably weak or even undetectable anisotropies are typically found for $sp_z$-type surface or quantum well states in systems with $C_{3v}$-symmetry, such as e.g. Cu(111) \cite{Reinert:01.03}, Bi$_2$Se$_3$(0001) \cite{Kuroda:10.8} or Ag/Si(111) \cite{Ogawa:12}. However, lowering the spatial symmetry and going to more directional orbitals can drastically enhance the anisotropy of band dispersions as demonstrated in Ref.~\onlinecite{Miyamoto:10.12} for the $d$-type surface state on W(110) with $C_{2v}$-symmetry.    

Our measurements directly image the quantized band structure and Fermi surface
of ferromagnetic Fe films on a small energy scale. The energy
separation of the observed QWS is $\sim$50~meV and thus of the same order as
temperature-induced broadening effects at $E_F$ at room temperature
(4$kT\approx$~100~meV). Magnetic quantum size effects originating from such
closely spaced QWS are expected to become significantly damped when approaching
room temperature because the discrete character of the near-$E_F$ electronic
structure becomes washed out \cite{himpsel:99.10,cinal:03}. Oscillations in the
magnetic anisotropy in the system Fe/Ag(1,1,10) were indeed found to vanish
above 200~K and our results, though not directly comparable due to the
different substrate material, provide an important hint on the origin of this
observation \cite{Wu:09.5}. As a strategy to enhance the energy spacing a
possibility is to go to lower films thicknesses or to use different ferromagnetic materials \cite{Bauer:11.10}.

In summary, the presented ARPES experiments reveal a quantized electronic fine structure close to the Fermi energy of 4--4.5~nm Fe(110) films being relevant for making use of quantum magnetic size effects in metallic multilayers. The observed QWS appear on a low energy scale in the meV regime indicating that quantization effects in ferromagnetic layers can occur on the same energy scale as classical many-body phenomena. The observed QWS do not follow a simple free-electron behavior but rather exhibit a strong anisotropy in their dispersion within the surface plane. Their effective mass changes by a factor of $\sim$50 between the $\overline{\Gamma \rm{H}}$- and $\overline{\Gamma \rm{N}}$-directions. Density functional theory calculations largely agree with the experimental findings. Based on our theoretical results we attribute the strong electronic anisotropy both to the intrinsic electronic structure of {\it bcc} Fe and to extrinsic influences caused by coupling to the W(110) substrate.

\section{Acknowledgment}
We would like to thank Markus Donath for fruitful discussions. This work was financially supported by the Deutsche Forschungsgemeinschaft through FOR1162 and project SFB-F41 (ViCoM) of the Austrian Science Fund. 

\bibliographystyle{apsrev}

\begin{thebibliography}{39}
\expandafter\ifx\csname natexlab\endcsname\relax\def\natexlab#1{#1}\fi
\expandafter\ifx\csname bibnamefont\endcsname\relax
  \def\bibnamefont#1{#1}\fi
\expandafter\ifx\csname bibfnamefont\endcsname\relax
  \def\bibfnamefont#1{#1}\fi
\expandafter\ifx\csname citenamefont\endcsname\relax
  \def\citenamefont#1{#1}\fi
\expandafter\ifx\csname url\endcsname\relax
  \def\url#1{\texttt{#1}}\fi
\expandafter\ifx\csname urlprefix\endcsname\relax\def\urlprefix{URL }\fi
\providecommand{\bibinfo}[2]{#2}
\providecommand{\eprint}[2][]{\url{#2}}

\bibitem[{\citenamefont{Cox et~al.}(1993)\citenamefont{Cox, Louderback, and
  Bloomfield}}]{Cox:93}
\bibinfo{author}{\bibfnamefont{A.~J.} \bibnamefont{Cox}},
  \bibinfo{author}{\bibfnamefont{J.~G.} \bibnamefont{Louderback}},
  \bibnamefont{and} \bibinfo{author}{\bibfnamefont{L.~A.}
  \bibnamefont{Bloomfield}}, \bibinfo{journal}{Phys. Rev. Lett.}
  \textbf{\bibinfo{volume}{71}}, \bibinfo{pages}{923} (\bibinfo{year}{1993}).

\bibitem[{\citenamefont{Chiang}(2000)}]{Chiang:00}
\bibinfo{author}{\bibfnamefont{T.-C.} \bibnamefont{Chiang}},
  \bibinfo{journal}{Surface Science Reports} \textbf{\bibinfo{volume}{39}},
  \bibinfo{pages}{181 } (\bibinfo{year}{2000}).

\bibitem[{\citenamefont{Wie{\ss}ner et~al.}(2012)\citenamefont{Wie{\ss}ner,
  Lastra, Ziroff, Forster, Puschnig, D\"ossel, M\"ullen, Sch\"oll, and
  Reinert}}]{wiessner:12}
\bibinfo{author}{\bibfnamefont{M.}~\bibnamefont{Wie{\ss}ner}},
  \bibinfo{author}{\bibfnamefont{N.~S.~R.} \bibnamefont{Lastra}},
  \bibinfo{author}{\bibfnamefont{J.}~\bibnamefont{Ziroff}},
  \bibinfo{author}{\bibfnamefont{F.}~\bibnamefont{Forster}},
  \bibinfo{author}{\bibfnamefont{P.}~\bibnamefont{Puschnig}},
  \bibinfo{author}{\bibfnamefont{L.}~\bibnamefont{D\"ossel}},
  \bibinfo{author}{\bibfnamefont{K.}~\bibnamefont{M\"ullen}},
  \bibinfo{author}{\bibfnamefont{A.}~\bibnamefont{Sch\"oll}}, \bibnamefont{and}
  \bibinfo{author}{\bibfnamefont{F.}~\bibnamefont{Reinert}},
  \bibinfo{journal}{New Journal of Physics} \textbf{\bibinfo{volume}{14}},
  \bibinfo{pages}{113008} (\bibinfo{year}{2012}).

\bibitem[{\citenamefont{K\"onig et~al.}(2007)\citenamefont{K\"onig, Wiedmann,
  Br\"une, Roth, Buhmann, Molenkamp, Qi, and Zhang}}]{Molenkamp:07}
\bibinfo{author}{\bibfnamefont{M.}~\bibnamefont{K\"onig}},
  \bibinfo{author}{\bibfnamefont{S.}~\bibnamefont{Wiedmann}},
  \bibinfo{author}{\bibfnamefont{C.}~\bibnamefont{Br\"une}},
  \bibinfo{author}{\bibfnamefont{A.}~\bibnamefont{Roth}},
  \bibinfo{author}{\bibfnamefont{H.}~\bibnamefont{Buhmann}},
  \bibinfo{author}{\bibfnamefont{L.~W.} \bibnamefont{Molenkamp}},
  \bibinfo{author}{\bibfnamefont{X.}~\bibnamefont{Qi}}, \bibnamefont{and}
  \bibinfo{author}{\bibfnamefont{S.}~\bibnamefont{Zhang}},
  \bibinfo{journal}{Science} \textbf{\bibinfo{volume}{318}},
  \bibinfo{pages}{766} (\bibinfo{year}{2007}).

\bibitem[{\citenamefont{Bode et~al.}(2007)\citenamefont{Bode, Heide, von
  Bergmann, Ferriani, Heinze, Bihlmayer, Kubetzka, Pietzsch, Bl\"ugel, and
  Wiesendanger}}]{Bode:07}
\bibinfo{author}{\bibfnamefont{M.}~\bibnamefont{Bode}},
  \bibinfo{author}{\bibfnamefont{M.}~\bibnamefont{Heide}},
  \bibinfo{author}{\bibfnamefont{K.}~\bibnamefont{von Bergmann}},
  \bibinfo{author}{\bibfnamefont{P.}~\bibnamefont{Ferriani}},
  \bibinfo{author}{\bibfnamefont{S.}~\bibnamefont{Heinze}},
  \bibinfo{author}{\bibfnamefont{G.}~\bibnamefont{Bihlmayer}},
  \bibinfo{author}{\bibfnamefont{A.}~\bibnamefont{Kubetzka}},
  \bibinfo{author}{\bibfnamefont{O.}~\bibnamefont{Pietzsch}},
  \bibinfo{author}{\bibfnamefont{S.}~\bibnamefont{Bl\"ugel}}, \bibnamefont{and}
  \bibinfo{author}{\bibfnamefont{R.}~\bibnamefont{Wiesendanger}},
  \bibinfo{journal}{Nature} \textbf{\bibinfo{volume}{447}},
  \bibinfo{pages}{190} (\bibinfo{year}{2007}).

\bibitem[{\citenamefont{Smith et~al.}(1994)\citenamefont{Smith, Brookes, Chang,
  and Johnson}}]{Smith:94.1}
\bibinfo{author}{\bibfnamefont{N.~V.} \bibnamefont{Smith}},
  \bibinfo{author}{\bibfnamefont{N.~B.} \bibnamefont{Brookes}},
  \bibinfo{author}{\bibfnamefont{Y.}~\bibnamefont{Chang}}, \bibnamefont{and}
  \bibinfo{author}{\bibfnamefont{P.~D.} \bibnamefont{Johnson}},
  \bibinfo{journal}{Phys. Rev. B} \textbf{\bibinfo{volume}{49}},
  \bibinfo{pages}{332} (\bibinfo{year}{1994}).

\bibitem[{\citenamefont{Guo et~al.}(2004)\citenamefont{Guo, Zhang, Bao, Han,
  Tang, Zhang, Zhu, Wang, Niu, Qiu et~al.}}]{Xue:04}
\bibinfo{author}{\bibfnamefont{Y.}~\bibnamefont{Guo}},
  \bibinfo{author}{\bibfnamefont{Y.-F.} \bibnamefont{Zhang}},
  \bibinfo{author}{\bibfnamefont{X.-Y.} \bibnamefont{Bao}},
  \bibinfo{author}{\bibfnamefont{T.-Z.} \bibnamefont{Han}},
  \bibinfo{author}{\bibfnamefont{Z.}~\bibnamefont{Tang}},
  \bibinfo{author}{\bibfnamefont{L.-X.} \bibnamefont{Zhang}},
  \bibinfo{author}{\bibfnamefont{W.-G.} \bibnamefont{Zhu}},
  \bibinfo{author}{\bibfnamefont{E.~G.} \bibnamefont{Wang}},
  \bibinfo{author}{\bibfnamefont{Q.}~\bibnamefont{Niu}},
  \bibinfo{author}{\bibfnamefont{Z.~Q.} \bibnamefont{Qiu}},
  \bibnamefont{et~al.}, \bibinfo{journal}{Science}
  \textbf{\bibinfo{volume}{306}}, \bibinfo{pages}{1915} (\bibinfo{year}{2004}).

\bibitem[{\citenamefont{Luh et~al.}(2002)\citenamefont{Luh, Miller, Paggel, and
  Chiang}}]{Chiang:02.7}
\bibinfo{author}{\bibfnamefont{D.-A.} \bibnamefont{Luh}},
  \bibinfo{author}{\bibfnamefont{T.}~\bibnamefont{Miller}},
  \bibinfo{author}{\bibfnamefont{J.~J.} \bibnamefont{Paggel}},
  \bibnamefont{and} \bibinfo{author}{\bibfnamefont{T.-C.}
  \bibnamefont{Chiang}}, \bibinfo{journal}{Phys. Rev. Lett.}
  \textbf{\bibinfo{volume}{88}}, \bibinfo{pages}{256802}
  (\bibinfo{year}{2002}).

\bibitem[{\citenamefont{Zhang et~al.}(2005)\citenamefont{Zhang, Jia, Han, Tang,
  Shen, Guo, Qiu, and Xue}}]{Xue:05}
\bibinfo{author}{\bibfnamefont{Y.-F.} \bibnamefont{Zhang}},
  \bibinfo{author}{\bibfnamefont{J.-F.} \bibnamefont{Jia}},
  \bibinfo{author}{\bibfnamefont{T.-Z.} \bibnamefont{Han}},
  \bibinfo{author}{\bibfnamefont{Z.}~\bibnamefont{Tang}},
  \bibinfo{author}{\bibfnamefont{Q.-T.} \bibnamefont{Shen}},
  \bibinfo{author}{\bibfnamefont{Y.}~\bibnamefont{Guo}},
  \bibinfo{author}{\bibfnamefont{Z.~Q.} \bibnamefont{Qiu}}, \bibnamefont{and}
  \bibinfo{author}{\bibfnamefont{Q.-K.} \bibnamefont{Xue}},
  \bibinfo{journal}{Phys. Rev. Lett.} \textbf{\bibinfo{volume}{95}},
  \bibinfo{pages}{096802} (\bibinfo{year}{2005}).

\bibitem[{\citenamefont{Bentmann et~al.}(2012)\citenamefont{Bentmann,
  Abdelouahed, Mulazzi, Henk, and Reinert}}]{Bentmann:12.05}
\bibinfo{author}{\bibfnamefont{H.}~\bibnamefont{Bentmann}},
  \bibinfo{author}{\bibfnamefont{S.}~\bibnamefont{Abdelouahed}},
  \bibinfo{author}{\bibfnamefont{M.}~\bibnamefont{Mulazzi}},
  \bibinfo{author}{\bibfnamefont{J.}~\bibnamefont{Henk}}, \bibnamefont{and}
  \bibinfo{author}{\bibfnamefont{F.}~\bibnamefont{Reinert}},
  \bibinfo{journal}{Phys. Rev. Lett.} \textbf{\bibinfo{volume}{108}},
  \bibinfo{pages}{196801} (\bibinfo{year}{2012}).

\bibitem[{\citenamefont{Li et~al.}(2009)\citenamefont{Li, Przybylski, Yildiz,
  Ma, and Wu}}]{Wu:09.5}
\bibinfo{author}{\bibfnamefont{J.}~\bibnamefont{Li}},
  \bibinfo{author}{\bibfnamefont{M.}~\bibnamefont{Przybylski}},
  \bibinfo{author}{\bibfnamefont{F.}~\bibnamefont{Yildiz}},
  \bibinfo{author}{\bibfnamefont{X.~D.} \bibnamefont{Ma}}, \bibnamefont{and}
  \bibinfo{author}{\bibfnamefont{Y.~Z.} \bibnamefont{Wu}},
  \bibinfo{journal}{Phys. Rev. Lett.} \textbf{\bibinfo{volume}{102}},
  \bibinfo{pages}{207206} (\bibinfo{year}{2009}).

\bibitem[{\citenamefont{Speer et~al.}(2006)\citenamefont{Speer, Tang, Miller,
  and Chiang}}]{speer:03.06}
\bibinfo{author}{\bibfnamefont{N.~J.} \bibnamefont{Speer}},
  \bibinfo{author}{\bibfnamefont{S.-J.} \bibnamefont{Tang}},
  \bibinfo{author}{\bibfnamefont{T.}~\bibnamefont{Miller}}, \bibnamefont{and}
  \bibinfo{author}{\bibfnamefont{T.-C.} \bibnamefont{Chiang}},
  \bibinfo{journal}{Science} \textbf{\bibinfo{volume}{314}},
  \bibinfo{pages}{804} (\bibinfo{year}{2006}).

\bibitem[{\citenamefont{Forster et~al.}(2011)\citenamefont{Forster, Gergert,
  Nuber, Bentmann, Huang, Gong, Zhang, and Reinert}}]{forster2011}
\bibinfo{author}{\bibfnamefont{F.}~\bibnamefont{Forster}},
  \bibinfo{author}{\bibfnamefont{E.}~\bibnamefont{Gergert}},
  \bibinfo{author}{\bibfnamefont{A.}~\bibnamefont{Nuber}},
  \bibinfo{author}{\bibfnamefont{H.}~\bibnamefont{Bentmann}},
  \bibinfo{author}{\bibfnamefont{L.}~\bibnamefont{Huang}},
  \bibinfo{author}{\bibfnamefont{X.~G.} \bibnamefont{Gong}},
  \bibinfo{author}{\bibfnamefont{Z.}~\bibnamefont{Zhang}}, \bibnamefont{and}
  \bibinfo{author}{\bibfnamefont{F.}~\bibnamefont{Reinert}},
  \bibinfo{journal}{Phys. Rev. B} \textbf{\bibinfo{volume}{84}},
  \bibinfo{pages}{075412} (\bibinfo{year}{2011}).

\bibitem[{\citenamefont{Reinert et~al.}(2000)\citenamefont{Reinert, Nicolay,
  Eltner, Ehm, Schmidt, H\"ufner, Probst, and Bucher}}]{reinert00}
\bibinfo{author}{\bibfnamefont{F.}~\bibnamefont{Reinert}},
  \bibinfo{author}{\bibfnamefont{G.}~\bibnamefont{Nicolay}},
  \bibinfo{author}{\bibfnamefont{B.}~\bibnamefont{Eltner}},
  \bibinfo{author}{\bibfnamefont{D.}~\bibnamefont{Ehm}},
  \bibinfo{author}{\bibfnamefont{S.}~\bibnamefont{Schmidt}},
  \bibinfo{author}{\bibfnamefont{S.}~\bibnamefont{H\"ufner}},
  \bibinfo{author}{\bibfnamefont{U.}~\bibnamefont{Probst}}, \bibnamefont{and}
  \bibinfo{author}{\bibfnamefont{E.}~\bibnamefont{Bucher}},
  \bibinfo{journal}{Phys. Rev. Lett.} \textbf{\bibinfo{volume}{85}},
  \bibinfo{pages}{3930} (\bibinfo{year}{2000}).

\bibitem[{\citenamefont{Vyalikh et~al.}(2008)\citenamefont{Vyalikh,
  Danzenb\"acher, Yaresko, Holder, Kucherenko, Laubschat, Krellner, Hossain,
  Geibel, Shi et~al.}}]{Laubschat:08}
\bibinfo{author}{\bibfnamefont{D.~V.} \bibnamefont{Vyalikh}},
  \bibinfo{author}{\bibfnamefont{S.}~\bibnamefont{Danzenb\"acher}},
  \bibinfo{author}{\bibfnamefont{A.~N.} \bibnamefont{Yaresko}},
  \bibinfo{author}{\bibfnamefont{M.}~\bibnamefont{Holder}},
  \bibinfo{author}{\bibfnamefont{Y.}~\bibnamefont{Kucherenko}},
  \bibinfo{author}{\bibfnamefont{C.}~\bibnamefont{Laubschat}},
  \bibinfo{author}{\bibfnamefont{C.}~\bibnamefont{Krellner}},
  \bibinfo{author}{\bibfnamefont{Z.}~\bibnamefont{Hossain}},
  \bibinfo{author}{\bibfnamefont{C.}~\bibnamefont{Geibel}},
  \bibinfo{author}{\bibfnamefont{M.}~\bibnamefont{Shi}}, \bibnamefont{et~al.},
  \bibinfo{journal}{Phys. Rev. Lett.} \textbf{\bibinfo{volume}{100}},
  \bibinfo{pages}{056402} (\bibinfo{year}{2008}).

\bibitem[{\citenamefont{Santander-Syro
  et~al.}(2009)\citenamefont{Santander-Syro, Klein, Boariu, Nuber, Lejay, and
  Reinert}}]{santander-syro:09}
\bibinfo{author}{\bibfnamefont{A.~F.} \bibnamefont{Santander-Syro}},
  \bibinfo{author}{\bibfnamefont{M.}~\bibnamefont{Klein}},
  \bibinfo{author}{\bibfnamefont{F.~L.} \bibnamefont{Boariu}},
  \bibinfo{author}{\bibfnamefont{A.}~\bibnamefont{Nuber}},
  \bibinfo{author}{\bibfnamefont{P.}~\bibnamefont{Lejay}}, \bibnamefont{and}
  \bibinfo{author}{\bibfnamefont{F.}~\bibnamefont{Reinert}},
  \bibinfo{journal}{Nature Physics} \textbf{\bibinfo{volume}{5}}
  (\bibinfo{year}{2009}).

\bibitem[{\citenamefont{Klein et~al.}(2008{\natexlab{a}})\citenamefont{Klein,
  Zur, Menzel, Schoenes, Doll, R\"oder, and Reinert}}]{Klein:08.07}
\bibinfo{author}{\bibfnamefont{M.}~\bibnamefont{Klein}},
  \bibinfo{author}{\bibfnamefont{D.}~\bibnamefont{Zur}},
  \bibinfo{author}{\bibfnamefont{D.}~\bibnamefont{Menzel}},
  \bibinfo{author}{\bibfnamefont{J.}~\bibnamefont{Schoenes}},
  \bibinfo{author}{\bibfnamefont{K.}~\bibnamefont{Doll}},
  \bibinfo{author}{\bibfnamefont{J.}~\bibnamefont{R\"oder}}, \bibnamefont{and}
  \bibinfo{author}{\bibfnamefont{F.}~\bibnamefont{Reinert}},
  \bibinfo{journal}{Phys. Rev. Lett.} \textbf{\bibinfo{volume}{101}},
  \bibinfo{pages}{046406} (\bibinfo{year}{2008}{\natexlab{a}}).

\bibitem[{\citenamefont{Zakeri et~al.}(2010)\citenamefont{Zakeri, Peixoto,
  Zhang, Prokop, and Kirschner}}]{Zakeri2010}
\bibinfo{author}{\bibfnamefont{K.}~\bibnamefont{Zakeri}},
  \bibinfo{author}{\bibfnamefont{T.}~\bibnamefont{Peixoto}},
  \bibinfo{author}{\bibfnamefont{Y.}~\bibnamefont{Zhang}},
  \bibinfo{author}{\bibfnamefont{J.}~\bibnamefont{Prokop}}, \bibnamefont{and}
  \bibinfo{author}{\bibfnamefont{J.}~\bibnamefont{Kirschner}},
  \bibinfo{journal}{Surface Science Letters} \textbf{\bibinfo{volume}{604}},
  \bibinfo{pages}{L1} (\bibinfo{year}{2010}).

\bibitem[{\citenamefont{Blaha et~al.}(2001)\citenamefont{Blaha, Schwarz,
  Madsen, Kvasnicka, and Luitz}}]{wien2k}
\bibinfo{author}{\bibfnamefont{P.}~\bibnamefont{Blaha}},
  \bibinfo{author}{\bibfnamefont{K.}~\bibnamefont{Schwarz}},
  \bibinfo{author}{\bibfnamefont{G.~K.~H.} \bibnamefont{Madsen}},
  \bibinfo{author}{\bibfnamefont{D.}~\bibnamefont{Kvasnicka}},
  \bibnamefont{and} \bibinfo{author}{\bibfnamefont{J.}~\bibnamefont{Luitz}},
  \emph{\bibinfo{title}{WIEN2k, An Augmented Plane Wave Plus Local Orbitals
  Program for Calculating Crystal Properties. ISBN 3-9501031-1-2}},
  \bibinfo{address}{Vienna University of Technology, Austria}
  (\bibinfo{year}{2001}).

\bibitem[{\citenamefont{Perdew et~al.}(1996)\citenamefont{Perdew, Burke, and
  Ernzerhof}}]{pbe}
\bibinfo{author}{\bibfnamefont{J.~P.}~\bibnamefont{Perdew}},
  \bibinfo{author}{\bibfnamefont{K.}~\bibnamefont{Burke}}, \bibnamefont{and}
  \bibinfo{author}{\bibfnamefont{M.}~\bibnamefont{Ernzerhof}},
  \bibinfo{journal}{Phys.\ Rev. Lett.} \textbf{\bibinfo{volume}{77}},
  \bibinfo{pages}{3865} (\bibinfo{year}{1996}).

\bibitem[{\citenamefont{Sch{\"a}fer et~al.}(2007)\citenamefont{Sch{\"a}fer,
  Hoinkis, Rotenberg, Blaha, and Claessen}}]{Schafer2007}
\bibinfo{author}{\bibfnamefont{J.}~\bibnamefont{Sch{\"a}fer}},
  \bibinfo{author}{\bibfnamefont{M.}~\bibnamefont{Hoinkis}},
  \bibinfo{author}{\bibfnamefont{E.}~\bibnamefont{Rotenberg}},
  \bibinfo{author}{\bibfnamefont{P.}~\bibnamefont{Blaha}}, \bibnamefont{and}
  \bibinfo{author}{\bibfnamefont{R.}~\bibnamefont{Claessen}},
  \bibinfo{journal}{Phys. Rev. B} \textbf{\bibinfo{volume}{75}},
  \bibinfo{pages}{092401} (\bibinfo{year}{2007}).

\bibitem[{\citenamefont{Klein et~al.}(2008{\natexlab{b}})\citenamefont{Klein,
  Zur, Menzel, Schoenes, Doll, R\"oder, and Reinert}}]{Klein:08}
\bibinfo{author}{\bibfnamefont{M.}~\bibnamefont{Klein}},
  \bibinfo{author}{\bibfnamefont{D.}~\bibnamefont{Zur}},
  \bibinfo{author}{\bibfnamefont{D.}~\bibnamefont{Menzel}},
  \bibinfo{author}{\bibfnamefont{J.}~\bibnamefont{Schoenes}},
  \bibinfo{author}{\bibfnamefont{K.}~\bibnamefont{Doll}},
  \bibinfo{author}{\bibfnamefont{J.}~\bibnamefont{R\"oder}}, \bibnamefont{and}
  \bibinfo{author}{\bibfnamefont{F.}~\bibnamefont{Reinert}},
  \bibinfo{journal}{Phys. Rev. Lett.} \textbf{\bibinfo{volume}{101}},
  \bibinfo{pages}{046406} (\bibinfo{year}{2008}{\natexlab{b}}).

\bibitem[{\citenamefont{Eiguren et~al.}(2002)\citenamefont{Eiguren, Hellsing,
  Reinert, Nicolay, Chulkov, Silkin, H\"ufner, and Echenique}}]{eiguren:02.01}
\bibinfo{author}{\bibfnamefont{A.}~\bibnamefont{Eiguren}},
  \bibinfo{author}{\bibfnamefont{B.}~\bibnamefont{Hellsing}},
  \bibinfo{author}{\bibfnamefont{F.}~\bibnamefont{Reinert}},
  \bibinfo{author}{\bibfnamefont{G.}~\bibnamefont{Nicolay}},
  \bibinfo{author}{\bibfnamefont{E.~V.} \bibnamefont{Chulkov}},
  \bibinfo{author}{\bibfnamefont{V.~M.} \bibnamefont{Silkin}},
  \bibinfo{author}{\bibfnamefont{S.}~\bibnamefont{H\"ufner}}, \bibnamefont{and}
  \bibinfo{author}{\bibfnamefont{P.~M.} \bibnamefont{Echenique}},
  \bibinfo{journal}{Phys. Rev. Lett.} \textbf{\bibinfo{volume}{88}},
  \bibinfo{pages}{066805} (\bibinfo{year}{2002}).

\bibitem[{\citenamefont{Reinert et~al.}(2003)\citenamefont{Reinert, Eltner,
  Nicolay, Ehm, Schmidt, and H\"ufner}}]{reinert:03.10}
\bibinfo{author}{\bibfnamefont{F.}~\bibnamefont{Reinert}},
  \bibinfo{author}{\bibfnamefont{B.}~\bibnamefont{Eltner}},
  \bibinfo{author}{\bibfnamefont{G.}~\bibnamefont{Nicolay}},
  \bibinfo{author}{\bibfnamefont{D.}~\bibnamefont{Ehm}},
  \bibinfo{author}{\bibfnamefont{S.}~\bibnamefont{Schmidt}}, \bibnamefont{and}
  \bibinfo{author}{\bibfnamefont{S.}~\bibnamefont{H\"ufner}},
  \bibinfo{journal}{Phys. Rev. Lett.} \textbf{\bibinfo{volume}{91}},
  \bibinfo{pages}{186406} (\bibinfo{year}{2003}).

\bibitem[{\citenamefont{Okuda et~al.}(2009)\citenamefont{Okuda, Takeichi, He,
  Harasawa, Kakizaki, and Matsuda}}]{Okuda:09.09}
\bibinfo{author}{\bibfnamefont{T.}~\bibnamefont{Okuda}},
  \bibinfo{author}{\bibfnamefont{Y.}~\bibnamefont{Takeichi}},
  \bibinfo{author}{\bibfnamefont{K.}~\bibnamefont{He}},
  \bibinfo{author}{\bibfnamefont{A.}~\bibnamefont{Harasawa}},
  \bibinfo{author}{\bibfnamefont{A.}~\bibnamefont{Kakizaki}}, \bibnamefont{and}
  \bibinfo{author}{\bibfnamefont{I.}~\bibnamefont{Matsuda}},
  \bibinfo{journal}{Phys. Rev. B} \textbf{\bibinfo{volume}{80}},
  \bibinfo{pages}{113409} (\bibinfo{year}{2009}).

\bibitem[{\citenamefont{Zhong et~al.}(2013)\citenamefont{Zhong, Q., and
  Held}}]{Zhong:13}
\bibinfo{author}{\bibfnamefont{Z.}~\bibnamefont{Zhong}},
  \bibinfo{author}{\bibfnamefont{Q.}~\bibnamefont{Zhang}}, \bibnamefont{and}
  \bibinfo{author}{\bibfnamefont{K.}~\bibnamefont{Held}},
  \bibinfo{journal}{Phys.\ Rev. B} \textbf{\bibinfo{volume}{88}},
  \bibinfo{pages}{125401} (\bibinfo{year}{2013}).

\bibitem[{\citenamefont{Assmann et~al.}(2013)\citenamefont{Assmann, Blaha,
  Laskowski, Held, Okamoto, and Sangiovanni}}]{Assmann:13}
\bibinfo{author}{\bibfnamefont{E.}~\bibnamefont{Assmann}},
  \bibinfo{author}{\bibfnamefont{P.}~\bibnamefont{Blaha}},
  \bibinfo{author}{\bibfnamefont{R.}~\bibnamefont{Laskowski}},
  \bibinfo{author}{\bibfnamefont{K.}~\bibnamefont{Held}},
  \bibinfo{author}{\bibfnamefont{S.}~\bibnamefont{Okamoto}}, \bibnamefont{and}
  \bibinfo{author}{\bibfnamefont{G.}~\bibnamefont{Sangiovanni}},
  \bibinfo{journal}{Phys.\ Rev. Lett.} \textbf{\bibinfo{volume}{110}},
  \bibinfo{pages}{078701} (\bibinfo{year}{2013}).

\bibitem[{\citenamefont{Mugarza et~al.}(2001)\citenamefont{Mugarza, Mascaraque,
  Perez-Dieste, Repain, Rousset, Garcia~de Abajo, and
  Ortega}}]{steppedgold}
\bibinfo{author}{\bibfnamefont{A.}~\bibnamefont{Mugarza}},
  \bibinfo{author}{\bibfnamefont{A.}~\bibnamefont{Mascaraque}},
  \bibinfo{author}{\bibfnamefont{V.}~\bibnamefont{Perez-Dieste}},
  \bibinfo{author}{\bibfnamefont{V.}~\bibnamefont{Repain}},
  \bibinfo{author}{\bibfnamefont{S.}~\bibnamefont{Rousset}},
  \bibinfo{author}{\bibfnamefont{F.~J.}~\bibnamefont{Garcia~de Abajo}},
  \bibnamefont{and} \bibinfo{author}{\bibfnamefont{J.~E.}~\bibnamefont{Ortega}},
  \bibinfo{journal}{Physical Review Letters} \textbf{\bibinfo{volume}{87}},
  \bibinfo{pages}{107601} (\bibinfo{year}{2001}).

\bibitem[{\citenamefont{Namba et~al.}(1996)\citenamefont{Namba, Nakanishi,
  Yamaguchi, Ohta, and Kuroda}}]{steppednickel2}
\bibinfo{author}{\bibfnamefont{H.}~\bibnamefont{Namba}},
  \bibinfo{author}{\bibfnamefont{N.}~\bibnamefont{Nakanishi}},
  \bibinfo{author}{\bibfnamefont{T.}~\bibnamefont{Yamaguchi}},
  \bibinfo{author}{\bibfnamefont{T.}~\bibnamefont{Ohta}}, \bibnamefont{and}
  \bibinfo{author}{\bibfnamefont{H.}~\bibnamefont{Kuroda}},
  \bibinfo{journal}{Surface Science} \textbf{\bibinfo{volume}{357-358}},
  \bibinfo{pages}{238} (\bibinfo{year}{1996}).

\bibitem[{\citenamefont{van~der Veen et~al.}(1981)\citenamefont{van~der Veen,
  Eastman, Bradshaw, and Holloway}}]{steppediridium}
\bibinfo{author}{\bibfnamefont{J.}~\bibnamefont{van~der Veen}},
  \bibinfo{author}{\bibfnamefont{D.}~\bibnamefont{Eastman}},
  \bibinfo{author}{\bibfnamefont{A.}~\bibnamefont{Bradshaw}}, \bibnamefont{and}
  \bibinfo{author}{\bibfnamefont{S.}~\bibnamefont{Holloway}},
  \bibinfo{journal}{Solid State Communications} \textbf{\bibinfo{volume}{39}},
  \bibinfo{pages}{1301} (\bibinfo{year}{1981}).

\bibitem[{\citenamefont{Huang et~al.}(2007)\citenamefont{Huang, Gong, Gergert,
  Forster, Bendounan, Reinert, and Zhang}}]{huang2007}
\bibinfo{author}{\bibfnamefont{L.}~\bibnamefont{Huang}},
  \bibinfo{author}{\bibfnamefont{X.~G.} \bibnamefont{Gong}},
  \bibinfo{author}{\bibfnamefont{E.}~\bibnamefont{Gergert}},
  \bibinfo{author}{\bibfnamefont{F.}~\bibnamefont{Forster}},
  \bibinfo{author}{\bibfnamefont{A.}~\bibnamefont{Bendounan}},
  \bibinfo{author}{\bibfnamefont{F.}~\bibnamefont{Reinert}}, \bibnamefont{and}
  \bibinfo{author}{\bibfnamefont{Z.}~\bibnamefont{Zhang}},
  \bibinfo{journal}{EPL} \textbf{\bibinfo{volume}{78}}, \bibinfo{pages}{57003}
  (\bibinfo{year}{2007}).

\bibitem[{\citenamefont{Miyamoto et~al.}(2012)\citenamefont{Miyamoto, Kimura,
  Okuda, Shimada, Iwasawa, Hayashi, Namatame, Taniguchi, and
  Donath}}]{Miyamoto:10.12}
\bibinfo{author}{\bibfnamefont{K.}~\bibnamefont{Miyamoto}},
  \bibinfo{author}{\bibfnamefont{A.}~\bibnamefont{Kimura}},
  \bibinfo{author}{\bibfnamefont{T.}~\bibnamefont{Okuda}},
  \bibinfo{author}{\bibfnamefont{K.}~\bibnamefont{Shimada}},
  \bibinfo{author}{\bibfnamefont{H.}~\bibnamefont{Iwasawa}},
  \bibinfo{author}{\bibfnamefont{H.}~\bibnamefont{Hayashi}},
  \bibinfo{author}{\bibfnamefont{H.}~\bibnamefont{Namatame}},
  \bibinfo{author}{\bibfnamefont{M.}~\bibnamefont{Taniguchi}},
  \bibnamefont{and} \bibinfo{author}{\bibfnamefont{M.}~\bibnamefont{Donath}},
  \bibinfo{journal}{Phys. Rev. B} \textbf{\bibinfo{volume}{86}},
  \bibinfo{pages}{161411} (\bibinfo{year}{2012}).

\bibitem[{\citenamefont{Crain et~al.}(2004)\citenamefont{Crain, McChesney,
  Zheng, Gallagher, Snijders, Bissen, Gundelach, Erwin, and
  Himpsel}}]{Crain:04}
\bibinfo{author}{\bibfnamefont{J.~N.} \bibnamefont{Crain}},
  \bibinfo{author}{\bibfnamefont{J.~L.} \bibnamefont{McChesney}},
  \bibinfo{author}{\bibfnamefont{F.}~\bibnamefont{Zheng}},
  \bibinfo{author}{\bibfnamefont{M.~C.} \bibnamefont{Gallagher}},
  \bibinfo{author}{\bibfnamefont{P.~C.} \bibnamefont{Snijders}},
  \bibinfo{author}{\bibfnamefont{M.}~\bibnamefont{Bissen}},
  \bibinfo{author}{\bibfnamefont{C.}~\bibnamefont{Gundelach}},
  \bibinfo{author}{\bibfnamefont{S.~C.} \bibnamefont{Erwin}}, \bibnamefont{and}
  \bibinfo{author}{\bibfnamefont{F.~J.} \bibnamefont{Himpsel}},
  \bibinfo{journal}{Phys. Rev. B} \textbf{\bibinfo{volume}{69}},
  \bibinfo{pages}{125401} (\bibinfo{year}{2004}).

\bibitem[{\citenamefont{Reinert et~al.}(2001)\citenamefont{Reinert, Nicolay,
  Schmidt, Ehm, and H\"ufner}}]{Reinert:01.03}
\bibinfo{author}{\bibfnamefont{F.}~\bibnamefont{Reinert}},
  \bibinfo{author}{\bibfnamefont{G.}~\bibnamefont{Nicolay}},
  \bibinfo{author}{\bibfnamefont{S.}~\bibnamefont{Schmidt}},
  \bibinfo{author}{\bibfnamefont{D.}~\bibnamefont{Ehm}}, \bibnamefont{and}
  \bibinfo{author}{\bibfnamefont{S.}~\bibnamefont{H\"ufner}},
  \bibinfo{journal}{Phys. Rev. B} \textbf{\bibinfo{volume}{63}},
  \bibinfo{pages}{115415} (\bibinfo{year}{2001}).

\bibitem[{\citenamefont{Kuroda et~al.}(2010)\citenamefont{Kuroda, Arita,
  Miyamoto, Ye, Jiang, Kimura, Krasovskii, Chulkov, Iwasawa, Okuda
  et~al.}}]{Kuroda:10.8}
\bibinfo{author}{\bibfnamefont{K.}~\bibnamefont{Kuroda}},
  \bibinfo{author}{\bibfnamefont{M.}~\bibnamefont{Arita}},
  \bibinfo{author}{\bibfnamefont{K.}~\bibnamefont{Miyamoto}},
  \bibinfo{author}{\bibfnamefont{M.}~\bibnamefont{Ye}},
  \bibinfo{author}{\bibfnamefont{J.}~\bibnamefont{Jiang}},
  \bibinfo{author}{\bibfnamefont{A.}~\bibnamefont{Kimura}},
  \bibinfo{author}{\bibfnamefont{E.~E.} \bibnamefont{Krasovskii}},
  \bibinfo{author}{\bibfnamefont{E.~V.} \bibnamefont{Chulkov}},
  \bibinfo{author}{\bibfnamefont{H.}~\bibnamefont{Iwasawa}},
  \bibinfo{author}{\bibfnamefont{T.}~\bibnamefont{Okuda}},
  \bibnamefont{et~al.}, \bibinfo{journal}{Phys. Rev. Lett.}
  \textbf{\bibinfo{volume}{105}}, \bibinfo{pages}{076802}
  (\bibinfo{year}{2010}).

\bibitem[{\citenamefont{Ogawa et~al.}(2012)\citenamefont{Ogawa, Sheverdyaeva,
  Moras, Topwal, Harasawa, Kobayashi, Carbone, and Matsuda}}]{Ogawa:12}
\bibinfo{author}{\bibfnamefont{M.}~\bibnamefont{Ogawa}},
  \bibinfo{author}{\bibfnamefont{P.~M.} \bibnamefont{Sheverdyaeva}},
  \bibinfo{author}{\bibfnamefont{P.}~\bibnamefont{Moras}},
  \bibinfo{author}{\bibfnamefont{D.}~\bibnamefont{Topwal}},
  \bibinfo{author}{\bibfnamefont{A.}~\bibnamefont{Harasawa}},
  \bibinfo{author}{\bibfnamefont{K.}~\bibnamefont{Kobayashi}},
  \bibinfo{author}{\bibfnamefont{C.}~\bibnamefont{Carbone}}, \bibnamefont{and}
  \bibinfo{author}{\bibfnamefont{I.}~\bibnamefont{Matsuda}},
  \bibinfo{journal}{Journal of Physics: Condensed Matter}
  \textbf{\bibinfo{volume}{24}}, \bibinfo{pages}{115501}
  (\bibinfo{year}{2012}).

\bibitem[{\citenamefont{Himpsel et~al.}(1999)\citenamefont{Himpsel, Altmann,
  Mankey, Ortega, and Petrovykh}}]{himpsel:99.10}
\bibinfo{author}{\bibfnamefont{F.}~\bibnamefont{Himpsel}},
  \bibinfo{author}{\bibfnamefont{K.}~\bibnamefont{Altmann}},
  \bibinfo{author}{\bibfnamefont{G.}~\bibnamefont{Mankey}},
  \bibinfo{author}{\bibfnamefont{J.}~\bibnamefont{Ortega}}, \bibnamefont{and}
  \bibinfo{author}{\bibfnamefont{D.}~\bibnamefont{Petrovykh}},
  \bibinfo{journal}{Journal of Magnetism and Magnetic Materials}
  \textbf{\bibinfo{volume}{200}}, \bibinfo{pages}{456} (\bibinfo{year}{1999}).

\bibitem[{\citenamefont{Cinal}(2003)}]{cinal:03}
\bibinfo{author}{\bibfnamefont{M.}~\bibnamefont{Cinal}},
  \bibinfo{journal}{Journal of Physics: Condensed Matter}
  \textbf{\bibinfo{volume}{15}}, \bibinfo{pages}{29} (\bibinfo{year}{2003}).

\bibitem[{\citenamefont{Bauer et~al.}(2011)\citenamefont{Bauer, Dabrowski,
  Przybylski, and Kirschner}}]{Bauer:11.10}
\bibinfo{author}{\bibfnamefont{U.}~\bibnamefont{Bauer}},
  \bibinfo{author}{\bibfnamefont{M.}~\bibnamefont{Dabrowski}},
  \bibinfo{author}{\bibfnamefont{M.}~\bibnamefont{Przybylski}},
  \bibnamefont{and}
  \bibinfo{author}{\bibfnamefont{J.}~\bibnamefont{Kirschner}},
  \bibinfo{journal}{Phys. Rev. B} \textbf{\bibinfo{volume}{84}},
  \bibinfo{pages}{144433} (\bibinfo{year}{2011}).

\end{thebibliography}

\end{document}